# PRESLA: An Original Device to Measure the Mechanical Interaction between Tongue and Teeth or Palate during Speech Production

Christophe Jeannin[1,3], Pascal Perrier[1], Yohan Payan[4], Brigitte Grosgogeat[2,3], André Dittmar[5], Claudine Géhin[5]

[1] ICP/GIPSA-lab, UMR CNRS 5216, Grenoble INP, France
[2] Laboratoire des Multimatériaux et Interface, UMR CNRS 5615, Université Lyon I, France
[3] Hospices Civils de Lyon, Service d'odontologie, Lyon, France
[4] TIMC-IMAG, UMR CNRS 5525, UJF, Grenoble, France
[5] Laboratoire des microsystèmes et microcapteurs biomédicaux, INSA, Lyon, France

E-mail: `Pascal.Perrier@gipsa-lab.inpg.fr`

**Abstract**

*An original experimental procedure is presented to measure the mechanical interaction between tongue and teeth and palate during speech production. It consists in using edentulous people as subjects and to insert pressure sensors in the structure of a replication of their dental prosthesis. This is assumed to induce no speech production perturbation for subjects who are used to speak with their prosthesis. Data collected from 4 subjects of French demonstrate the usability of the system.*

## 1 Introduction

It has been hypothesized in the literature ([1], [2]) that the speech motor control system could make use of external mechanical structures such as the palate or the teeth to induce specific strains in the tongue and to increase speech movement accuracy during the production of palatal and alveopalatal stops and fricatives and of high vowels. Quantitative measurements of the mechanical pressure exerted by the tongue on the teeth and the palate could provide an interesting basis to quantitatively assess this hypothesis. A number of experimental set-ups have been developed in the past years ([3], [4]) to measure this pressure. The limits of these techniques, beside the inherent complexity of their calibration, lie in the fact that they actually induce slight perturbations onto speech production, because they modify the geometry of the vocal tract. In this paper, we present an original experimental procedure, PRESLA ([5], [6]) that has not this drawback, because its design has been specifically adapted to a peculiar population of subjects, namely edentulous subjects who wear a complete dental prosthesis.

## 2 Materials and method

### 2.1 The *PRESLA* system

The dental prosthesis consists of a complete artificial denture and of an artificial palate. Artificial teeth are similar in shape and size to natural teeth (Figure 1). The false palate lowers the palatal vault and modifies its shape. However since our subjects are used to wear the prosthesis, it can be assumed that it does not correspond to a perturbation any longer. For the pressure measurements; a strain gauge sensor is inserted into a duplicate of the complete denture (Figure1, left panel). Since the artificial palate carrying a complete artificial denture must be at least 3mm thick to avoid breakage during mastication, both the pressure sensor and the wires connecting it with the connector outside of the mouth can be easily inserted in the prosthesis. This ensures that as compared to the normal prosthesis, our experimental setup does not induce any additional change of the geometry of the oral cavity. Thus, it can be hypothesized that for edentulous subjects who are used to speak with their prosthesis, our experimental device does not induce any perturbation of the speech production.

The position of the transducer in the complete denture varies with the sound that is analyzed. It is determined by a palatographic recording (Figure1, right panel) in order to locate the position of the contacts and to measure tongue pressure as accurately as possible for each sound under investigation.

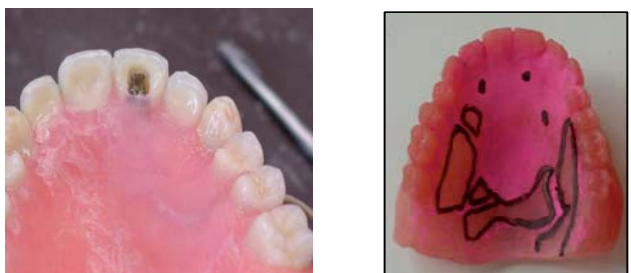

Figure 1: *Artificial complete denture.*
*Left panel: strain gauge sensor inserted in a tooth.*
*Right panel: Palatographic recording for /t/.*

The transducers are handmade and their locations in the complete denture are different for each patient. Hence, it is necessary to calibrate each sensor in the prosthesis for each subject specifically. In this aim, a calibration device has been developed, called Dried Water Column (DWC) (Figure 2, left panel). It consists of a long vertical tube in which water can be filled in. At the lower tip of the water column, a deformable water container made of latex (Figure 2 - right panel) is in contact with the pressure sensor that is in place in the complete denture. This set-up allows giving a fair account of the visco-elastic properties of tongue soft tissues and of their strain when they are in contact with rigid bodies.

The container has a convex shape due to the water pressure that is accurately controlled by the height of the water in the column. The mechanical pressure is thus applied to the sensor as long as the latex container is not too much stiffened (Figure 2, left panels). If this condition is not strictly respected, the mechanical pressure is also a function of the surface tension of the latex due to Laplace forces, and the electrical signal delivered by strain gauge sensor is no more a true measure of the water pressure.

In appropriate conditions of experimentation, an accurate conversion of the voltage of the electrical signal delivered by the sensor into pressure values is possible.

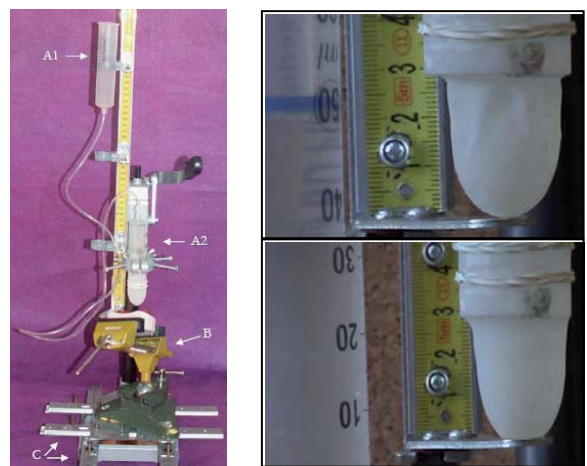

Figure 2: *PRESLA'S calibration system.*
*Left panel: General view; water pressure is controlled by container A1 height; the latex container forms the tip of container A2, which is connected to A1 via a flexible tube.*
*Left panel: Examples of shape of the latex container allowing a true measurement of the water pressure*

The relation between sensor signal and water pressure was found to be linear and without any hysteresis effect in the pressure range going from 0 to 80 mm of water. In this range of pressure, the voltage of the sensor signal is the order of a few microvolts. Hence, measurement was done thanks to the use of a Wheastone bridge that was supplied by a 2V DC voltage. DC supplier's (Vishay 2120A) output voltage presents very small instabilities, which are of the same order of magnitude than the measured signal but at a much higher frequency. Hence, the signal delivered by the sensor was band-pass filtered before being analyzed (0dB in the range [3Hz–70Hz].

### 2.2 Method

Measurements were carried out for 4 subjects (GRO, LEN, THE, BRE) who were all native speakers of French. All the subjects were wearing their complete denture for months and were used to speak with it. The true replica of the complete denture is then assumed to correspond to their current normal speaking conditions. The corpus consists on the one hand of many repetitions of the syllable /ta/ (condition henceforth called *reiterated speech*, and, on the other hand, of repetitions of the short meaningful French sentence "Toto a tété sa tétine" /totoatetesatetin/ (henceforth called *meaningful* *speech*). In both cases subjects were asked to speak at a normal speaking rate.

In addition to these recordings under normal condition, recordings were made under perturbed conditions, in order to explore the influence of orosensory and auditory feedbacks tongue/palate interaction in speech production. In this aim, these feedbacks were perturbed respectively via a lingual anaesthesia and noisy headphones.

Recordings were made in a normal dentist room of Lyon's University Hospital during 4 different sessions at a one month interval. In each session, the normal condition was recorded first, followed by the condition with noisy auditory feedback and by the condition with lingual anaesthesia.

## 3 Results

### 3.1 Normal conditions

Figure 3 presents a typical example of the filtered signal delivered by the strain gauge sensor for speaker GRO during the production of reiterated speech. This example is representative for the four subjects, in terms of shape, timing and amplitude.

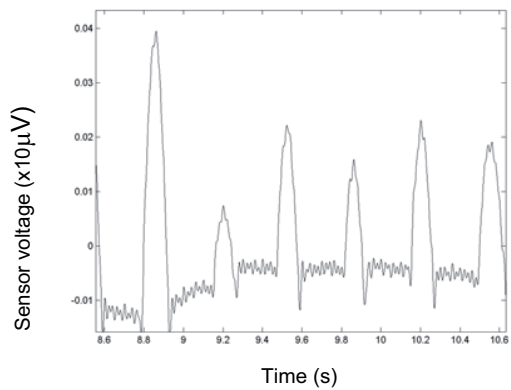

*Figure 3: Example of the sensor signal during reiterated speech (Speaker GRO).*

Noticeable intra-speaker variability is observed for the peak of pressure during the production of /t/: it ranges from 5cmH$_2$0 to 40cmH$_2$0. It varies with the repetition, with the position in the sentence (for meaningful speech, see below) and with the session. Figure 4 shows this /t/ pressure for speaker LEN during a complete recording session. It was observed that the pressure and its variability is similar for *reiterated speech* (left part of the graph) and for *meaningful speech* (right part of the graph). It should also be noted that the inter-speaker variability does not exceed the intra-speaker one.

Neel et al. ([7]) suggested that speech production would generate tongue/palate pressure in the range of 25% of the maximal pressure (Pmax) that the tongue can exert against the palate. In this context, the measurements from our subjects would correspond to Pmax values ranging from almost 20cmH$_2$0 up to around 160cmH$_2$0. This is in agreement with data provided in the literature ([8], [9]).

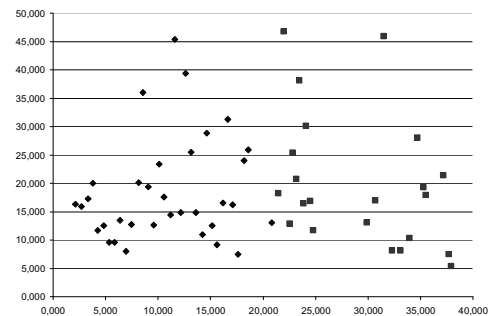

*Figure 4: Maximal pressure measured during /t/ occlusion for speaker LEN during the complete 3$^{rd}$ recording session. Reiterated speech is on the left (lozenges); meaningful speech is on the right (squares). X-axis: Time in s; Y-axis: Pressure in cmH$_2$0*

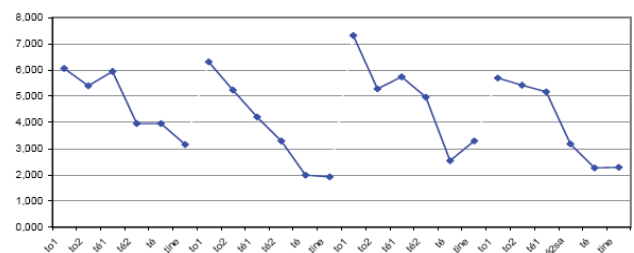

*Figure 5: Variation of maximal pressure (in cmH$_2$0) measured during /t/ occlusion for 4 repetitions of the meaningful sentence (speaker THE). Each measure corresponds to a syllable /t/-vowel within a sequence*

In meaningful speech a decrease of the pressure is systematically observed from the beginning to the end of each sentence (Figure 5). A similar decrease is observed for the intensity of the acoustic signal. This phenomenon is consistent with the declension prosodic effect in French (see for example [10].

### 3.1 Perturbed conditions

In perturbed conditions the maximal pressure measured during /t/ occlusion, in *reiterated* and in *meaningful speech* was compared to the measurements obtained in normal condition. For the 4 subjects no significant differences could be observed between the three conditions (see Table 1 summarizing the results for subject LEN).

The results obtained in Anaesthesia condition confirm our hypothesis that our subjects were used to speak with their complete denture. The damping of the orosensory feedback does not prevent them to produce alveolar stops as under normal condition. However, the results in Noise condition do not support the widely accepted assumption that speaking in noise is associated with an increase of articulatory effort (Lombard effect). This assumption was recently supported by Garnier et al. [11], who observed for bilabial stop consonants that under noisy conditions speakers tend to press their lips stronger against each other. Further studies seem necessary to confirm our findings.

*Table 1: Average maximal pressure (cmH$_2$0) during /t/.occlusion and its standard deviation (in parentheses) in meaningful speech measured for subject LEN across sessions and conditions*

| Condition | Session 1 | Session 2 |
|---|---|---|
| Normal | 23 (7.4) | 13.6 (6.2) |
| Noisy headphones | 17.7 (7.5) | 11.1 (4.1) |
| Anesthesia | 13.9 (7.5) | *No measure* |

| Condition | Session 3 | Session 4 |
|---|---|---|
| Normal | 19.9 (11.7) | 11.8 (9.4) |
| Noisy headphones | 18.1 (7.5) | 12.1 (4.9) |
| Anesthesia | 20.8 (10.6) | 9.6 (8.7) |

### 4. Conclusion

The range of pressure values found for the 4 speakers, its consistency across speakers, together with the pressure decrease observed within each sentence, demonstrate the usability of PRESLA. Further studies will aim at finding the origin of the large intra-speaker pressure variability and at clarifying and interpreting our observations in noisy feedback condition.

### References

[1] Fuchs, S., Perrier, P., Geng, C. & Mooshammer, C. (2006). What role does the palate play in speech motor control? Insights from tongue kinematics for German alveolar obstruents. In J. Harrington & M. Tabain (eds). *Speech Production: Models, Phonetic Processes, and Techniques* (pp. 148-164). Psychology Press: Sydney.

[2] Hamlet, S.L. and Stone, M. (1978). Compensatory alveolar consonant production induced by wearing a dental prosthesis. *Journal of Phonetics, 6*, 227-248.

[3] Kelly S., Mai A., Manley G. and McLean C.(2000). Electropalatography and the linguagraph system. *Medical Engineering & Physics, 22( 1),* 47-58

[4] Wakumoto, M., Masaki, S., Honda, K. & Ohue, T. (1998). A pressure sensitive palatography : Application of new pressure sensitive sheet for measuring tongue-palatal contact pressure. *Proceedings of the 5$^{th}$ Int. Conf. on Spoken Language Processing, Vol. 7,* 3151 – 3154.

[5] Jeannin, C., Perrier, P., Payan, Y., Dittmar, A. & Grosgogeat, B. (2005). A non-invasive device to measure mechanical interaction between tongue, palate and teeth during speech production. P*roceedings of MAVEBA2005*

[6] Jeannin, C., Perrier, P., Payan, Y., Dittmar, A., & Grosgogeat, B. (2008). Tongue pressure recordings during speech using complete denture *Materials Science and Engineering C, 28 (5-6)*, 835-841

[7] Neel AT, Palmer, PM, Sprouls, G, Morrison, L (2006) Tongue strength and speech intelligibility in oculopharyngeal muscular dystrophy. *J Med Speech-Language pathology, 14(4)*, 273-277.

[8] Hewitt A, Hind J, Kays S, Nicosia M, Doyle J, Tompkins W, Gangnon R, Robbins J.(2007). Standardized Instrument for Lingual Pressure Measurement., *Dysphagia*, *23(1)*, 16-25

[9] Matsumura, M, Yamasaki, H, Tsuji, R, Niikawa, T, Hara, H, Tachimura, T, Wada, T. (2000) Measurement of palatolingual contact pressure and tongue force using a force-sensor mounted on a palatal plate. In *Proceedings of ICSLP2000*, pp 893-896

[10] Tabain M. & Perrier P. (2005) Articulation and acoustics of /i/ in preboundarry position in French. *Journal of Phonetics*, *33*, 77–100

[11] Garnier, M, Henrich, N, Dubois, D, Poitevineau, J, Polack, JD. (2006) Peut-on considérer l'effet Lombard comme un phénomène linéaire en fonction du niveau de bruit ? In *Proceedings of the Congrès Français d'Acoustique 2006* (pp. 1053-1056), Paris, France